\documentclass[12pt]{article}
\usepackage{amsfonts}
\usepackage{eurosym}
\usepackage{amssymb}
\usepackage{amsmath}
\usepackage{cite}

\newtheorem{theorem}{Theorem}

\input{tcilatex}
\begin{document}

\title{Classification of the Lie and Noether symmetries for the Klein-Gordon
equation in Anisotropic Cosmology}
\author{Andronikos Paliathanasis\thanks{%
Email: anpaliat@phys.uoa.gr} \\
{\ {Institute of Systems Science, Durban University of Technology }}\\
{\ {PO Box 1334, Durban 4000, Republic of South Africa}} \\
Departamento de Matem\'{a}ticas, Universidad Cat\'{o}lica del Norte,\\
Avda. Angamos 0610, Casilla 1280 Antofagasta, Chile
}
\maketitle

\begin{abstract}
We carried out the detailed group classification of the potential in
Klein-Gordon equation in anisotropic Riemannian manifolds. Specifically, we
consider the Klein-Gordon equations for the four-dimensional anisotropic and
homogeneous spacetimes of Bianchi I, Bianchi\ III and Bianchi V. We derive
all the closed-form expressions for the potential function where the
equation admits Lie and Noether symmetries. We apply previous results which
connect the Lie symmetries\ of the differential equation with the
collineations of the Riemannian space which defines the Laplace operator,
and we solve the classification problem in a systematic way. Keywords: Lie
symmetries; Klein-Gordon; Anisotropic Spacetimes; Noether symmetries;
Conformal Killing Vectors
\end{abstract}

\section{Introduction}

\label{sec1}

A systematic approach for the study of nonlinear differential equations is
the Lie symmetry analysis \cite{ibra,Bluman,Stephani,olver}. The novelty of
the Lie symmetry approach is that through a systematic approach the
existence of invariant transformations can be determined. The latter can be
used to simplify the given differential equation with the use of similarity
transformations. Under the application of similarity transformations in a
given differential equation, we derive a new differential equation with less
independent variables. Furthermore, conservation laws can construct which
are essential for the study of the properties for the given differential
equation \cite{ibra}. Lie symmetries have been applied for the study of
nonlinear differential equations in all areas of applied mathematics \cite%
{mm1,mm2,mm3,mm4,mm5,mm6,mm7,mm8,mm9}.

A systematic approach for the construction and the determination of
conservation laws for differential equations was established by E. Noether.
In Noether's famous work of 1918 \cite{noe}, Noether showed that some of the
Lie symmetries were related to symmetries of the variational principle. For
each symmetry of the variation integral, Noether derived an exact formula
for the derivation of the conservation law. That very simple method for the
construction of conservation laws is very important in physical science and
in other theories of applied mathematics.

In General Relativity the natural space is of four-dimensions described by a
Riemannian manifold. In this work we investigate the Lie and Noether
symmetries for the Klein-Gordon equation in anisotropic homogeneous
geometries. Anisotropic homogeneous spacetimes are of special interest
because they can describe the very early period of the universe, that is,
before the inflationary era where anisotropies played an important role in
the evolution of the physical variables. There is a plethora of studies in
literature where symmetry analysis has been applied for the
classification of the geodesic equations \cite{sm1,sm2,sm3,sm4}, the wave
equation \cite{sm5,sm6} in curved spaces and the gravitational field
equations \cite{sm7,sm8,sm9,sm10}.

There has been investigated a relation between the symmetries of some
differential equations of special interest and the collineations of the
background geometry which provides the related differential operators.
Indeed, the Lie symmetries of the geodesic equations in Riemannian manifolds
are constructed by the elements of the projective group of the background
spacetime \cite{sp1,sp2}. The latter relation is true if we consider the
existence of a force term in the field equations \cite{sp3}. For the Noether
symmetries of the geodesic Lagrangian these are derived by the elements of
the homothetic group of the background geometry \cite{sp2}. That geometric
results are extended and for higher-order symmetries, see for instance the
discussion in \cite{sp4} and references therein. As far as the case of
partial differential equations is concerned, the symmetries of the Poisson
equation are constructed by the elements of the conformal algebra of the
Riemann metric which defines the Laplace operator \cite{sp6}. Hence, it is
clear that in order to solve the classification problem of our study we
should present in detail the classification of the conformal algebra for the
homogeneous and anisotropic spacetimes of our consideration. The structure
of the paper is as follows.

In Section \ref{sec2} we present in detail the theory of infinitesimal
transformation and the definitions of basic motions in Riemannian manifolds.
Moreover, we present the classification of \ the Killing symmetries, the
Homothetic vector and the proper Conformal Killing vector for the Bianchi I,
Bianchi III and Bianchi V spacetimes. In Section \ref{sec3} we present the
basic elements of the theory of differential equations. For the Poisson and
the Klein-Gordon equation we recover previous results which show how the Lie
symmetries are constructed directly from the Conformal Killing vectors of
the background geometry. Moreover, a similar result is also presented and
for the Noether symmetries of the Klein-Gordon equation. The classification
problem of our study is solved in Section \ref{sec4}. We present all the
functional forms of the potential function for the Klein-Gordon equation
where nontrivial symmetry vectors exist for the Klein-Gordon equation.
Finally, in Section \ref{sec6} we summarise our results.

\section{Infinitesimal Transformations and Motions of Riemannian spaces}

\label{sec2}

Assume the Riemannian manifold $V^{n}$, $\dim V^{n}=n,$ with metric tensor $%
g_{\mu \nu }$. Consider now the one-parameter point transformation defined
by the parametric equation $\bar{x}^{\mu }=\bar{x}^{\mu }\left( x^{\nu
},\varepsilon \right) ~$which defines a group orbit through the point $%
P\left( x^{\mu },0\right) $. Thus, the tangent vector at the point $P$ is
given by the following expression%
\begin{equation}
X=\frac{\partial \bar{x}^{\mu }}{\partial \varepsilon }|_{\varepsilon
\rightarrow 0}\partial _{x^{\mu }}|_{P}.  \label{pt.01}
\end{equation}

$X$ is the generator vector of the infinitesimal transformation near the
point $P$
\begin{equation}
\bar{x}^{\mu }=x^{\mu }+\varepsilon \xi ^{\mu }\left( x^{\nu }\right) ,
\label{pt.02}
\end{equation}%
in which $\xi ^{\mu }=\frac{\partial \bar{x}^{\mu }}{\partial \varepsilon }%
|_{\varepsilon \rightarrow 0}.$

Let $F\left( x^{\mu }\right) $ be a function in the Riemannian manifold
defined at the point $P$. \ Hence, under the action of the one-parameter
point transformation (\ref{pt.01}) the function reads $\bar{F}\left( \bar{x}%
^{\mu }\right) $.

By definition, function $F$ is invariant under the action of the one
parameter point transformation (\ref{pt.01}) if and only if it has the same
value/expression before and after the transformation. That is, $\bar{F}%
\left( \bar{x}^{\mu }\right) =F\left( x^{\mu }\right) $ or equivalently $%
\bar{F}\left( \bar{x}^{\mu }\right) =0$ when$~F\left( x^{\mu }\right) =0\,$.
The latter definition is described by the mathematical expression with the
use of the infinitessimal generator%
\begin{equation}
X\left( F\right) =0.  \label{pt.03}
\end{equation}%
equivalently $\xi ^{\mu }\frac{\partial F}{\partial x^{\mu }}=0$.

Expression (\ref{pt.03}) is the Lie symmetry condition for a function $%
F\left( x^{\mu }\right) $ to be invariant under the action of an
one-parameter point transformation in the base manifold. If condition (\ref%
{pt.03})\ is true for a specific vector field $X$, then $X$ is a Lie
symmetry vector for the function $F\left( x^{\mu }\right) $.

Consider now $\Omega ^{\mu }\left( x^{\nu }\right) $ to be a geometric
object with the generic transformation rule \cite{yano} $\bar{\Omega}^{\mu
\prime }=\Phi ^{\mu }\left( \Omega ^{\nu },x^{\nu },\bar{x}^{\nu }\right) .$%
When $\Omega ^{\mu }\left( x^{\nu }\right) $ is a linear homogeneous
geometric object the transformation rule reads \cite{yano} $\Phi ^{\mu
}\left( \Omega ^{\nu },x^{\nu },\bar{x}^{\nu }\right) =J_{\lambda }^{\mu
}\left( x^{\nu },\bar{x}^{\nu }\right) \Omega ^{\lambda }.$Where $J_{\lambda
}^{\mu }\left( x^{\nu },\bar{x}^{\nu }\right) $ is the Jacobian matrix for
the one-parameter point transformation with generator (\ref{pt.01}), that is
$J_{\lambda }^{\mu }=\frac{\partial \bar{x}^{\mu }}{\partial x^{\nu }}$.

Similar, with the definition of functions, the a geometric object $\Omega
^{\mu }\left( x^{\nu }\right) ~$is invariant under a one parameter point
transformation (\ref{pt.03}) if and only if $\bar{\Omega}^{\mu }\left( \bar{x%
}^{\nu }\right) =\Omega ^{\mu }\left( x^{\nu }\right) $ or
\begin{equation}
\mathcal{L}_{X}\Omega ^{\mu }\left( x^{\nu }\right) =0,  \label{pt.06}
\end{equation}%
where $\mathcal{L}_{X}$ is the Lie derivative with respect to the vector
field $X$. In the case where $\Omega ^{\mu }\left( x^{\nu }\right) \,$is a
function for the Lie derivative it holds $\mathcal{L}_{X}\Omega \equiv
X\left( \Omega \right) $.

A more generalized concept of the Lie symmetries for geometric objects are
summarized in the context of collineations. Consider now that for the
geometric object $\Omega $, the following expression holds%
\begin{equation}
\mathcal{L}_{X}\Omega =\Psi   \label{pt.07}
\end{equation}%
where $\Psi $ is a tensor field and it has the same components and
symmetries of the indices with $\Omega .~$If condition (\ref{pt.07}) is true
,$X$ is called a collineation for the geometric object $\Omega ,$\ then, the
type of collineations is being defined by tensor field $\Psi $.

The metric tensor $g_{\mu \nu }$ of the Riemannian manifold is a linear
homogeneous geometric object with definition for the Lie derivative
\begin{equation}
\mathcal{L}_{X}g_{\mu \nu }=X_{\left( \mu ;\nu \right) }
\end{equation}%
where $;$ denotes covariant derivative with respect to the Levi-Civita
connection.

For the metric tensor, the concept of collineations is expressed as
\begin{equation}
\mathcal{L}_{X}g_{\mu \nu }=2\psi g_{\mu \nu }+2H_{\mu \nu }
\end{equation}%
where $\psi $ is the conformal function and $H_{\mu \nu }$ is a symmetric
traceless tensor, i.e. $H_{~\mu }^{\mu }=0$. The most important
collineations for the metric tensor are the motions with $H_{\mu \nu }=0$.
These are the Killing vectors, the Homothetic vectors and the Conformal
killing vectors.

The generator (\ref{pt.01}) of the infinitesimal transformation (\ref{pt.02}%
) is called a Killing vector field (KV) for the Riemann space $V^{n},$~if
and only if the metric tensor is invariant under the action of the
transformation, that is,
\begin{equation}
\mathcal{L}_{X}g_{\mu \nu }=0.
\end{equation}

Moreover, the infinitesimal generator $X$ is a Conformal Killing vector
(CKV) for the Riemann space $V^{n}$ if there exists a function $\psi \left(
x^{\mu }\right) $ such that
\begin{equation}
\mathcal{L}_{X}g_{\mu \nu }=2\psi g_{\mu \nu }
\end{equation}%
where $\psi =\frac{1}{n}X_{;\mu }^{\mu }$.

An important class of collineations is when $\psi $ is a constant, then the
CKV becomes a Homothetic Killing Vector (HV). Moreover, when $\psi _{;\mu
\nu }=0$, the vector field $X$ is a special CKV (sp. CKV) for the Riemann
manifold. Indeed, when $\psi =0$, the CKV is also a KV. With the term proper
CKV we shall refer to CKVs which are not HVs or KVs.

The KVs, the HV and the CKVs form Lie algebras which are known as Killing
algebra $\left( G_{KV}\right) $, Homothetic algebra $\left( G_{HV}\right) ~$%
and Conformal Killing algebra $\left( G_{CV}\right) $. When for the
dimensional of the Riemannian manifold $V^{n}$ holds ~$n\geq 2,$ then $G_{KV}
$ is a subalgebra of $G_{HV}$ and the latter is a subalgebra of the
Conformal Killing algebra, that is $G_{KV}$ $\subseteq G_{HV}\subseteq G_{CV}
$. For any Riemannian manifold, there exists at most one proper Homothetic
vector. \ Moreover, the maximum dimensional Killing algebra is of $\frac{1}{2%
}n\left( n+1\right) $ and the maximum Conformal Killing algebra is of $\frac{%
1}{2}\left( n+1\right) \left( n+2\right) $ dimension.

Point transformations with a KV generator \ have the property to keep
invariant the length and the angles of autoparallels, unlike of the
homothetic vector where the angles remain invariant and the length is scaled
with a constant parameter. However, in the case where the point
transformation is generated by a CKV only the angles of autoparallels remain
invariant.

The existence of collineations for the metric tensor is essential for the
nature of the physical space which is described by Riemannian geometry.
Indeed, our universe in large scales is described by the Friedmann--Lema%
\^{\i}tre--Robertson--Walker line element which has a maximal symmetric
three-dimensional hypersurface. An important class of exact solutions in
General Relativity are the self-similar spacetimes. This family of solutions
has the main property to map to itself after an appropriate scale of the
dependent or independent variables, thus a proper HV exists. Self-similar
solutions of exact spacetimes describe the asymptotic behaviour of the most
general solution of the gravitational theory \cite{as1,as2}. Spacetimes with
proper\ CKV are also of special interest; more details can be found in \cite%
{bbokexactsolutions}. \

\subsection{CKVs of Anisotropic spacetimes}

The Bianchi spacetimes describe anisotropic homogeneous cosmologies and they
are of special interest, because they can describe the very early stage of
the evolution of universe. In this family of spacetimes the line element of
the metric tensor is foliated along the time axis, with three dimensional
homogeneous hypersurfaces. The classification problem of all three
dimensional real Lie algebras was solved by Bianchi and it has shown that
there are nine Lie algebras. Thus, there are nine Bianchi models according
to the admitted Killing algebra of the three-dimensional homogeneous
hypersurface.

The generic line element for the Bianchi model is%
\begin{equation}
ds^{2}=-N^{2}\left( t\right) dt^{2}+A^{2}(t)(\omega
_{1})^{2}+B^{2}(t)(\omega _{2})^{2}+C^{2}(t)(\omega _{3})^{2}\,\text{.}
\label{ln.01}
\end{equation}%
where $\omega _{i}$, $i=1,2,3$, are basic one-forms and $N\left( t\right)
,~A(t)$, $B(t)$, $C(t)$ are functions which depend only on the time
parameter, see \cite{ryan}. In this study we are interested in the Bianchi
I, Bianchi III and Bianchi V spacetimes. These spacetimes in terms of the
coordinate expressions are diagonal.

Indeed, for these spacetimes the $1-$forms are
\begin{eqnarray*}
\text{Bianchi I} &:&\omega _{1}=dx~,~\omega _{2}=dy~,~\omega _{3}=dz~, \\
\text{Bianchi III} &\text{:}&\omega _{1}=dx~,~\omega _{2}=dy~,~\omega
_{3}=e^{-x}dz~, \\
\text{Bianchi V} &\text{:}&\omega _{1}=dx~,~\omega _{2}=e^{x}dy~,~\omega
_{3}=e^{x}dz~.
\end{eqnarray*}

The Killing algebras of the Bianchi spacetimes are presented in \cite{ryan}.
However, the proper CKVs for the Bianchi I, Bianchi III and Bianchi V
spacetimes have been derived before in \cite{mt1,mt2}.

The Bianchi I spacetime admits proper CKV when the metric tensor provides
the line element%
\begin{equation}
ds^{2}=C^{2}(t)\left[ -dt^{2}+e^{-\frac{2}{c}t}dx^{2}+e^{-\frac{2\alpha _{1}%
}{c}t}dy^{2}+dz^{2}\right]  \label{ln.02}
\end{equation}%
with proper CKV the vector field%
\begin{equation}
X_{1}=c\partial _{t}+x\partial _{x}+\alpha _{1}y\partial _{y}
\end{equation}%
and conformal factor $\psi (X_{1})=c\left( \ln \left\vert C\right\vert
\right) _{,t}$. When $C\left( t\right) =e^{\psi _{0}t}$, $X_{1}$ reduces to
a proper HKV, while for $C\left( t\right) =C_{0}$, $X_{1}$ is a KV.

Moreover, when the line element is of the form
\begin{equation}
ds^{2}=C^{2}(t)\left[ -dt^{2}+t^{2\frac{\alpha _{2}-1}{\alpha _{2}}%
}dx^{2}+t^{2\frac{\alpha _{2}-\alpha _{1}}{\alpha _{2}}}dy^{2}+dz^{2}\right]
\label{ln.03}
\end{equation}%
the resulting CKV is

\begin{equation}
X_{2}=\alpha _{2}t\partial _{t}+x\partial _{x}+\alpha _{1}y\partial
_{y}+\alpha _{2}z\partial _{z}
\end{equation}%
with corresponding conformal factor $\psi (X_{2})=\alpha _{2}\left[ 1+t(\ln
\left\vert C\right\vert )_{,t}\right] $. Indeed, when $C\left( t\right)
=t^{\psi _{0}-1}$, $X_{2}$ is a proper HKV, while when $C\left( t\right)
=t^{-1}$, $X_{2}$ is reduced to a KV.

The Bianchi III spacetime admits a proper CKV when
\begin{equation}
ds^{2}=A^{2}\left( t\right) \left[ e^{m\lambda t}\left(
-dt^{2}+dx^{2}\right) +e^{m\left( \lambda -1\right) t}dy^{2}+e^{-2x}dz^{2}%
\right] .  \label{ln.04}
\end{equation}%
where now the corresponding vector field is
\begin{equation}
X_{3}=\frac{2}{m}\partial _{t}+y\partial _{y}+\lambda z\partial _{z}
\end{equation}%
and $\psi _{(III)}(X_{3})=\frac{2}{m}\frac{A_{,t}}{A}+\lambda $. Indeed,
when $A\left( t\right) =e^{A_{0}t}$, the vector field is reduced to a HV,
and for $A\left( t\right) =e^{-\frac{\lambda }{2}mx}$, is reduced to a KV.

Finally, for the family of Bianchi V spacetimes it follows that the line
element
\begin{equation}
ds^{2}=A^{2}\left( t\right) \left[ e^{m\lambda t}\left(
-dt^{2}+dx^{2}\right) +e^{2x}\left( e^{m(\lambda -1)t}dy^{2}+dz^{2}\right) %
\right]  \label{ln.05}
\end{equation}%
admits as proper CKV the vector field $X_{3}$ with the same conformal factor
as before.

Recall that in the following we shall not investigate the case where the
spacetimes reduce to locally rotational spaces or the scale factors are
constant functions.

\section{Symmetries of differential equations}

\label{sec3}

In terms of geometry a differential equation can be considered as a function
$H=H(x^{\nu },u^{A},u_{,\mu }^{A},u_{,\mu \nu }^{A})$ in the space $%
B=B\left( x^{\nu },u^{A},u_{,\mu }^{A},u_{,\mu \nu }^{A}\right) $, $%
u^{A}=u^{A}\left( x^{\mu }\right) $ denote the dependent variables, $x^{\mu
} $ are the independent variables and $u_{,\mu }^{A}=\frac{\partial u^{A}}{%
\partial u^{\mu }}$.

Assume now the infinitesimal transformation in the base manifold of the
differential equation $H$,
\begin{align}
\bar{x}^{\mu }& =x^{\mu }+\varepsilon \xi ^{\mu }(x^{\nu },u^{B})~,
\label{pr.01} \\
\bar{u}^{A}& =\bar{u}^{A}+\varepsilon \eta ^{A}(x^{\nu },u^{B})~,
\label{pr.02}
\end{align}%
with vector generator
\begin{equation}
\mathbf{X}=\xi ^{\mu }(x^{\nu },u^{B})\partial _{x^{\mu }}+\eta ^{A}(x^{\nu
},u^{B})\partial _{u^{A}}~.  \label{pr.03}
\end{equation}

Similarly to the case of functions, the geometric vector field\ $X\mathbf{\ }
$is a Lie symmetry of $H=H(x^{\nu },u^{A},u_{,\mu }^{A},u_{,\mu \nu }^{A})$
if and only if the following is true \cite{Stephani,Bluman}
\begin{equation}
X^{[2]}(H)=0,  \label{pr.04}
\end{equation}%
in which $X^{[2]}$ is the second extension of the vector field $X$ in the
space$~B=B\left( x^{\nu },u^{A},u_{,\mu }^{A},u_{,\mu \nu }^{A}\right) $
defined as
\begin{equation}
X^{[2]}=X+\eta _{\mu }^{A}\partial _{u_{\mu }^{A}}+\eta _{\mu \nu
}^{A}\partial _{u_{\mu \nu }^{A}}  \label{pr.05}
\end{equation}%
in which
\begin{equation}
\eta _{\mu }^{A}=D_{\mu }\eta ^{A}-D_{\mu }\xi ^{\nu }u_{,\nu }^{A}~,
\label{pr.06}
\end{equation}%
and
\begin{equation*}
\eta _{\mu \nu }^{A}=D_{\nu }\eta _{\mu }^{A}-D_{\nu }D_{\mu }\xi ^{\kappa
}u_{,\kappa }^{A}~,
\end{equation*}%
where $D_{\mu }~$is the total derivative.

A straightforward application of the Lie symmetries for a given differential
equation is the construction of invariant functions by deriving the
characteristic functions. The characteristic functions can be used to define
similarity transformations which can be used to reduce the number of the
indepedent variables in the case of partial differential equations.

The invariants are determined by the solution of the following Lagrangian
system.%
\begin{equation}
\frac{dx^{\mu }}{\xi ^{\mu }}=\frac{du^{A}}{\eta ^{A}}=\frac{du_{\mu }^{A}}{%
\eta _{\mu }^{A}}=\frac{du_{\mu \nu }^{A}}{\eta _{\mu \nu }^{A}}.
\end{equation}

In the case where the differential equation $H$ follows from a variational
principle with Lagrangian function $L=L(x^{\mu },u^{A},u_{,\mu }^{A})$ such
as $H\equiv \mathbf{E}\left( L\right) =0$, where $\mathbf{E}$ is the Euler
operator. The Lie point symmetry $\mathbf{X}$ of the DE\ $H\ \ $is a Noether
point symmetry of $H,\ $if and only if the following condition is satisfied%
\begin{equation}
\mathbf{X}^{[1]}L+LD_{i}\xi ^{i}=D_{i}A^{i}\left( x^{k},u^{C}\right) ,~
\label{pr.08}
\end{equation}%
where $\mathbf{X}^{[1]}$ is the first prolongation of $\mathbf{X,}$ and $%
A^{\mu }$ is a vector field which should be determined. Condition (\ref%
{pr.08}) is Noether's second theorem. The second theorem of Noether states
that for any vector field $X$ where condition (\ref{pr.08}) is true the
following function is a conservation law
\begin{equation}
I^{\mu }=\xi ^{\nu }\left( u_{\nu }^{A}\frac{\partial L}{\partial u_{\mu
}^{A}}-\delta _{\nu }^{\mu }L\right) -\eta ^{A}\frac{\partial L}{\partial
u_{\mu }^{A}}+A^{\mu }~,  \label{pr.09}
\end{equation}%
that is, $D_{\mu }I^{\mu }=0,~$\cite{Stephani,Bluman}.

\subsection{Poisson equation}

Let~$\Delta $ be the Laplace operator in the Riemannian manifold $V^{n}$,%
\begin{equation*}
\Delta =\frac{1}{\sqrt{g}}\frac{\partial }{\partial x^{\mu }}\left( \sqrt{g}%
g^{\mu \nu }\frac{\partial }{\partial x^{\nu }}\right) ,
\end{equation*}%
then the Poisson equation reads
\begin{equation}
\Delta u=f\left( x^{\mu },u\right) ,
\end{equation}%
or equivalently
\begin{equation}
g^{\mu \nu }u_{\mu \nu }-\Gamma ^{\mu }u_{\mu }=f\left( x^{\nu },u\right) ,
\label{PE.2}
\end{equation}%
where $\Gamma ^{\mu }=\Gamma _{\nu \kappa }^{\mu }g^{\nu \kappa }$ and $%
\Gamma _{\nu \kappa }^{\mu }$ are the Christoffel symbols for the
Levi-Civita connection of the metric tensor $g_{\mu \nu }$.

The Lie symmetry analysis for the Poisson equation when $f=f\left( u\right) $
have been given in \cite{Bozhkov}, and for $f=f\left( x^{i},u\right) $ are
presented in \cite{sp6}. Indeed, the Lie (point) symmetries for the Poisson
equation are related to the elements of the conformal algebra for the
Riemannian manifold as described in the following.

\begin{theorem}
\label{Theor}The Lie symmetries for the Poisson equation are constructed by
the generic CKV of the background metric tensor $g_{\mu \nu }$ of the
Riemannian manifold $V^{n}$:

a) For $n>2$,~the Lie symmetry vector has the generic form
\begin{equation}
X=\xi ^{\mu }\left( x^{\nu }\right) \partial _{\mu }+\left( \frac{2-n}{2}%
\psi \left( x^{\nu }\right) u+a_{0}u+b\left( x^{\nu }\right) \right)
\partial _{u},  \label{PE.7}
\end{equation}%
where $\xi ^{i}\left( x^{\nu }\right) $ is a CKV of the Riemannian manifold
with conformal factor $\psi \left( x^{\nu }\right) $ and the following
condition holds%
\begin{equation}
\frac{2-n}{2}\Delta \psi u+g^{\mu \nu }b_{\mu ;\nu }-\xi ^{\nu }f_{,\nu }-%
\frac{2-n}{2}\psi uf_{,u}-\frac{n+2}{2}\psi f-bf_{,u}=0,  \label{PE.8}
\end{equation}

b) For $n=2$, the generic Lie symmetry vector is%
\begin{equation}
X=\xi ^{\mu }\left( x^{\nu }\right) \partial _{\mu }+\left( a_{0}u+b\left(
x^{\nu }\right) \right) \partial _{u},
\end{equation}%
where $\xi ^{\mu }$ is a CKV and the following conditions are satisfied%
\begin{equation}
g^{\mu \nu }b_{;\mu \nu }-\xi ^{\nu }f_{,\nu }-a_{0}uf_{,u}+\left(
a_{0}-2\psi \right) f-bf_{,u}=0,
\end{equation}%
that is, the function $b$ is solution of the Laplace equation.
\end{theorem}

A special case of the Poisson equation is the Klein-Gordon equation with $%
f\left( x^{\nu },u\right) =V\left( x^{\nu }\right) u$, that is,
\begin{equation}
\Delta u-V\left( x^{\nu }\right) u=0,  \label{PE.10}
\end{equation}%
where $V(x^{\mu })$ is the potential function. For the Lie symmetries of the
Klein-Gordon equation it follows

\begin{theorem}
\label{KG}For the Klein Gordon equation (\ref{PE.10}) the Lie symmetries are
constructed by the elements of the conformal algebra of the Riemannian
manifold:

a) for $n>2$,~the generic symmetry vector is expressed as\newline
\begin{equation}
X=\xi ^{\mu }\left( x^{\nu }\right) \partial _{\mu }+\left( \frac{2-n}{2}%
\psi \left( x^{\nu }\right) u+a_{0}u+b\left( x^{\nu }\right) \right)
\partial _{u},  \label{PE.13}
\end{equation}%
where now $\xi ^{\mu }$ is a CKV with conformal factor $\psi \left( x^{\nu
}\right) ,$ $b\left( x^{\nu }\right) $ solves equation (\ref{PE.10}) with
constraint condition%
\begin{equation}
\xi ^{\nu }V_{,\nu }+2\psi V-\frac{2-n}{2}\Delta \psi =0,  \label{PE.14}
\end{equation}%
b) for $n=2$, the generic symmetry vector is written
\begin{equation}
X=\xi ^{\mu }\left( x^{\nu }\right) \partial _{\mu }+\left( a_{0}u+b\left(
x^{\nu }\right) \right) \partial _{u},
\end{equation}%
where $\xi ^{\mu }$ is a CKV with conformal factor $\psi \left( x^{\nu
}\right) ,$ $b\left( x^{\nu }\right) $ solves equation (\ref{PE.10}) with
constrain%
\begin{equation}
\xi ^{\nu }V_{,\nu }+2\psi V=0.
\end{equation}
\end{theorem}

The Klein-Gordon equation (\ref{PE.10}) can be reproduced by the variation
of the Lagrangian function%
\begin{equation}
L\left( x^{\nu },u,u_{,\nu }\right) =\frac{1}{2}\sqrt{\left\vert
g\right\vert }g^{\mu \nu }u_{,\mu }u_{,\nu }+\frac{1}{2}\sqrt{\left\vert
g\right\vert }V\left( x^{\nu }\right) u^{2}.  \label{p.01}
\end{equation}%
Therefore, for the Noether symmetries of the Klein-Gordon Lagrangian (\ref%
{p.01}) the following Theorem holds.

\begin{theorem}
\label{kgnoether}The Lie point symmetries of the Klein Gordon equation (\ref%
{PE.10}) are generated from the elements of the conformal algebra of the
Riemannian manifold, where the generic Noether symmetry is of the form%
\begin{equation*}
X_{N}=\xi ^{\mu }\left( x^{\nu }\right) \partial _{\mu }+\left( \frac{2-n}{2}%
\psi \left( x^{\nu }\right) u\right) \partial _{u},
\end{equation*}%
where the corresponding vector $A_{\mu }=\frac{2-n}{4}\sqrt{\left\vert
g\right\vert }\psi _{,\mu }\left( x^{\nu }\right) u^{2}$, in which $\xi
^{\mu }\left( x^{\nu }\right) $ is a CKV with conformal factor $\psi \left(
x^{\nu }\right) .$ The constraint equation is
\begin{equation}
\xi ^{\nu }V_{,\nu }+2\psi V-\frac{2-n}{2}\Delta \psi =0.
\end{equation}
\end{theorem}

We remark that for the Klein-Gordon equation all the non-trivial Lie
symmetries are also Noether symmetries. The resulting conservation law is of
the form%
\begin{equation}
I^{\mu }=\sqrt{\left\vert g\right\vert }\left( \left( \frac{1}{2}g^{\kappa
\nu }u_{,\kappa }u_{,\nu }-\frac{1}{2}V\left( x^{\nu }\right) u^{2}\right)
\xi ^{\mu }-\eta \frac{1}{2}g^{\mu \nu }u_{,\nu }+\frac{2-n}{4}\psi _{,\mu
}\left( x^{\nu }\right) u^{2}\right) .  \label{p.02}
\end{equation}

\section{Klein-Gordon equation in anisotropic geometries}

\label{sec4}

We proceed with the solution of the classification problem for the potential
function $V\left( x^{\mu }\right) $ for the Klein-Gordon equation (\ref%
{PE.10}) in the case of anisotropic cosmologies where the Klein-Gordon
equation admits non-trivial Lie symmetries. \ The trivial symmetries are the
vector fields $X_{u}=u\partial _{u}$, $X_{b}=b\partial _{u}$ which exist for
any potential function $V\left( x^{\nu }\right) $.

\subsection{Bianchi I}

In a Bianchi I spacetime, the Klein-Gordon equation is written%
\begin{equation}
\left( -\frac{u_{,tt}}{N^{2}}+\frac{u_{,xx}}{A^{2}}+\frac{u_{,yy}}{B^{2}}+%
\frac{u_{,zz}}{C^{2}}\right) +\frac{1}{N^{2}}\left( \frac{N_{.t}}{N}u_{,t}-%
\frac{A_{,t}}{A}u_{,x}-\frac{C_{,t}}{C}u_{,y}-\frac{C_{,t}}{C}u_{,z}\right) -%
\frac{V\left( t,x,y,z\right) }{A^{2}B^{2}C}u^{2}=0.  \label{kg.01}
\end{equation}

For arbitrary function forms of the scale factors the background space
admits the three KVs, $\xi _{I}^{1}=\partial _{x}~,~\xi _{I}^{2}=\partial
_{y}$ and $\xi _{I}^{3}=\partial _{z}$.

Hence, from Theorem \ref{KG} it follows that: $\ $(i) $\xi _{I}^{1}$, is a
Lie symmetry when $V_{i}^{I}\left( t,x,y,z\right) =V\left( t,y,z\right) $%
;~(ii) $\xi _{I}^{2}$ is a Lie symmetry when $V\left( t,x,y,z\right)
=V_{ii}^{I}\left( t,x,z\right) $; (iii) $\xi _{I}^{3}$ is a Lie symmetry
when $V\left( t,x,y,z\right) =V_{iii}^{I}\left( t,x,y\right) $; and (iv) $%
\alpha \xi _{I}^{1}+\beta \xi _{I}^{2}+\gamma \xi _{I}^{3}$ is a Lie
symmetry when $V_{iv}^{I}\left( t,x,y,z\right) =V\left( t,y-\frac{\beta }{%
\alpha }x,z-\frac{\gamma }{\alpha }x\right) $.

In the special case where the line element is that of (\ref{ln.02}) the CKV $%
X_{1}$ produces the Lie symmetry vector $X=X_{1}+\left( -2c\left( \ln
\left\vert C\right\vert \right) _{,t}u\right) \partial _{u}$ for the
Klein-Gordon equation, if and only if
\begin{equation}
V_{v}^{I}\left( t,x,y,z\right) =\frac{cC_{,tt}-\left( \alpha _{1}+1\right)
C_{,t}}{cC^{3}}+\frac{1}{C^{2}}U\left( xe^{-\frac{t}{c}},ye^{-\frac{a_{1}}{c}%
t},z\right) .
\end{equation}

Similarly, the vector field $X=X_{1}+\alpha \xi _{I}^{1}+\beta \xi
_{I}^{2}+\gamma \xi _{I}^{3}+\left( -2c\left( \ln \left\vert C\right\vert
\right) _{,t}u\right) \partial _{u}$ is a Lie symmetry for the Klein-Gordon
equation in a Bianchi I spacetime with line element (\ref{ln.02}) if and
only if
\begin{equation}
V_{vi}^{I}\left( t,x,y,z\right) =\frac{cC_{,tt}-\left( \alpha _{1}+1\right)
C_{,t}}{cC^{3}}+\frac{1}{C^{2}}U\left( \left( x+\alpha \right) e^{-\frac{t}{c%
}},\left( y+\frac{\beta }{a_{1}}\right) e^{-\frac{a_{1}}{c}t},z-\frac{\gamma
}{c}\tau \right) .
\end{equation}

On the other hand, for the Bianchi I line element (\ref{ln.03}) the CKV $%
X_{2}$ is the generator of the Lie symmetry vector $X=X_{2}-\alpha _{2}\left[
1+t(\ln \left\vert C\right\vert )_{,t}\right] u\partial _{u}$ for the
Klein-Gordon equation (\ref{kg.01}) for the potential function

\begin{equation}
V_{vii}^{I}\left( t,x,y,z\right) =\frac{1}{t^{2}C^{2}}\left( U\left( xt^{-%
\frac{1}{a_{2}}},yt^{-\frac{a_{1}}{a_{2}}},\frac{z}{t}\right) +F\left(
t\right) \right)
\end{equation}%
where
\begin{eqnarray}
F\left( t\right) &=&-\frac{1}{a_{2}C^{2}}\left(
a_{2}t^{2}CC_{,ttt}+tC_{,tt}\left( a_{2}C_{,t}+C\left( a_{1}-4a_{2}+1\right)
\right) \right) +  \notag \\
&&-\frac{1}{a_{2}C^{2}}\left( C_{,t}\left( 1+a_{1}-2a_{2}\right) \left(
C-tC_{,t}\right) \right) \text{\thinspace }.
\end{eqnarray}

Moreover, the vector field $X=X_{2}+\alpha \xi _{I}^{1}+\beta \xi
_{I}^{2}+\gamma \xi _{I}^{3}-\alpha _{2}\left[ 1+t(\ln \left\vert
C\right\vert )_{,t}\right] \partial _{u}$ is a Lie symmetry for the
Klein-Gordon equation when%
\begin{equation}
V_{viii}^{I}\left( t,x,y,z\right) =\frac{1}{t^{2}C^{2}}\left( U\left( \left(
x+\alpha \right) t^{-\frac{1}{a_{2}}},\left( y+\frac{\beta }{a_{1}}\right)
t^{-\frac{a_{1}}{a_{2}}},\left( z+\frac{\gamma }{a_{2}}\right) t^{-1}\right)
+F\left( t\right) \right) .
\end{equation}

\subsubsection{Invariant functions}

Let us now determine the invariant functions which correspond to each
admitted Lie point symmetry. The invariant functions can be used to
determine similarity transformations whenever they are applied the number of
dependent variables of the Klein-Gordon equation is reduced.

For the vector field $\xi _{I}^{1}$, the invariant functions are $\left\{
t,y,z,u\right\} $. Similarly, for the vector field $\xi _{I}^{2}$ we
determine the Lie invariants $\left\{ t,x,z,u\right\} $. Moreover, for $\xi
_{I}^{3}$ the Lie invariants are $\left\{ t,x,y,u\right\} $, while for the
vector field $\alpha \xi _{I}^{1}+\beta \xi _{I}^{2}+\gamma \xi _{I}^{3}$,
the Lie invariants are $\left\{ t,y-\frac{\beta }{\alpha }x,z-\frac{\gamma }{%
\alpha }x,u\right\} $.

Furthermore, for the potential function $V_{v}^{I}$ where $X=X_{1}+\left(
-2c\left( \ln \left\vert C\right\vert \right) _{,t}u\right) \partial _{u}$
is a Lie symmetry, the resulting Lie invariants are calculated $\left\{ xe^{-%
\frac{t}{c}},ye^{-\frac{a_{1}}{c}t},z,uC\left( t\right) ^{2}\right\} $. For
the potential $V_{vi}^{I}$, the Lie invariants are $\left\{ \left( x+\alpha
\right) e^{-\frac{t}{c}},\left( y+\frac{\beta }{a_{1}}\right) e^{-\frac{a_{1}%
}{c}t},z-\frac{\gamma }{c}\tau ,uC\left( t\right) ^{2}\right\} $. In a
similar way, for the potential functions $V_{vii}^{I}\left( t,x,y,z\right) $
and $V_{viii}^{I}\left( t,x,y,z\right) $ the admitted Lie invariants are $%
\left\{ xt^{-\frac{1}{a_{2}}},yt^{-\frac{a_{1}}{a_{2}}},\frac{z}{t}%
,ut^{2}C\left( t\right) ^{2}\right\} $and $\left\{ \left( x+\alpha \right)
t^{-\frac{1}{a_{2}}},\left( y+\frac{\beta }{a_{1}}\right) t^{-\frac{a_{1}}{%
a_{2}}},\left( z+\frac{\gamma }{a_{2}}\right) t^{-1},ut^{2}C\left( t\right)
^{2}\right\} $ provided by the Lie symmetries $X_{2}-\alpha _{2}\left[
1+t(\ln \left\vert C\right\vert )_{,t}\right] u\partial _{u}$ and $%
X_{2}+\alpha \xi _{I}^{1}+\beta \xi _{I}^{2}+\gamma \xi _{I}^{3}-\alpha _{2}%
\left[ 1+t(\ln \left\vert C\right\vert )_{,t}\right] \partial _{u}$
respectively.

\subsubsection{Conservation laws}

We apply Noether's theorem and expression (\ref{p.02}) hence the resulting
conservation laws related to the admitted Lie symmetries for the
Klein-Gordon equation (\ref{kg.01}) are
\begin{equation}
I^{x}\left( \xi _{I}^{1}\right) =\frac{NABC}{2}\left( \left( \left( -\frac{1%
}{N^{2}}u_{,t}^{2}+\frac{1}{A^{2}}u_{,x}^{2}+\frac{1}{B^{2}}u_{,y}^{2}+\frac{%
1}{C^{2}}u_{,z}^{2}\right) -V_{i}^{I}\left( t,x,y,z\right) u^{2}\right)
\right) .
\end{equation}%
\begin{equation}
I^{t}\left( \xi _{I}^{1}\right) =0~,~I^{y}\left( \xi _{I}^{1}\right) =0\text{
and }I^{z}\left( \xi _{I}^{1}\right) =0.
\end{equation}%
For the vector field $\xi _{I}^{2}$ there exists the conservation law%
\begin{equation}
I^{y}\left( \xi _{I}^{2}\right) =\frac{NABC}{2}\left( \left( \left( -\frac{1%
}{N^{2}}u_{,t}^{2}+\frac{1}{A^{2}}u_{,x}^{2}+\frac{1}{B^{2}}u_{,y}^{2}+\frac{%
1}{C^{2}}u_{,z}^{2}\right) -V_{ii}^{I}\left( t,x,y,z\right) u^{2}\right)
\right) .
\end{equation}%
\begin{equation}
I^{t}\left( \xi _{I}^{2}\right) =0~,~I^{x}\left( \xi _{I}^{2}\right) =0\text{
and }I^{z}\left( \xi _{I}^{2}\right) =0.
\end{equation}%
While for the vector field $\xi _{I}^{3}$ the resulting Noetherian
conservation law is
\begin{equation}
I^{z}\left( \xi _{I}^{3}\right) =\frac{NABC}{2}\left( \left( \left( -\frac{1%
}{N^{2}}u_{,t}^{2}+\frac{1}{A^{2}}u_{,x}^{2}+\frac{1}{B^{2}}u_{,y}^{2}+\frac{%
1}{C^{2}}u_{,z}^{2}\right) -V_{iii}^{I}\left( t,x,y,z\right) u^{2}\right)
\right) .
\end{equation}%
\begin{equation}
I^{t}\left( \xi _{I}^{3}\right) =0~,~I^{x}\left( \xi _{I}^{3}\right) =0\text{
and }I^{y}\left( \xi _{I}^{3}\right) =0.
\end{equation}

For the generic vector field $\alpha \xi _{I}^{1}+\beta \xi _{I}^{2}+\gamma
\xi _{I}^{3}$ we calculate the conservation law%
\begin{equation}
I^{x}\left( \alpha \xi _{I}^{1}+\beta \xi _{I}^{2}+\gamma \xi
_{I}^{3}\right) =\alpha \frac{NABC}{2}\left( \left( \left( -\frac{1}{N^{2}}%
u_{,t}^{2}+\frac{1}{A^{2}}u_{,x}^{2}+\frac{1}{B^{2}}u_{,y}^{2}+\frac{1}{C^{2}%
}u_{,z}^{2}\right) -V_{iv}^{I}\left( t,x,y,z\right) u^{2}\right) \right) ,
\end{equation}%
\begin{equation}
I^{y}\left( \alpha \xi _{I}^{1}+\beta \xi _{I}^{2}+\gamma \xi
_{I}^{3}\right) =\beta \frac{NABC}{2}\left( \left( \left( -\frac{1}{N^{2}}%
u_{,t}^{2}+\frac{1}{A^{2}}u_{,x}^{2}+\frac{1}{B^{2}}u_{,y}^{2}+\frac{1}{C^{2}%
}u_{,z}^{2}\right) -V_{iv}^{I}\left( t,x,y,z\right) u^{2}\right) \right) ,
\end{equation}%
\begin{equation}
I^{z}\left( \alpha \xi _{I}^{1}+\beta \xi _{I}^{2}+\gamma \xi
_{I}^{3}\right) =\gamma \frac{NABC}{2}\left( \left( \left( -\frac{1}{N^{2}}%
u_{,t}^{2}+\frac{1}{A^{2}}u_{,x}^{2}+\frac{1}{B^{2}}u_{,y}^{2}+\frac{1}{C^{2}%
}u_{,z}^{2}\right) -V_{iv}^{I}\left( t,x,y,z\right) u^{2}\right) \right)
\end{equation}%
\begin{equation}
I^{t}\left( \alpha \xi _{I}^{1}+\beta \xi _{I}^{2}+\gamma \xi
_{I}^{3}\right) =0~.
\end{equation}

For the potential $V_{v}^{I}\left( t,x,y,z\right) $ there exists the
conservation law%
\begin{equation}
I^{t}\left( X_{1}\right) =c\frac{e^{-\frac{t\left( 1+\alpha _{1}\right) }{c}%
}C^{4}}{2}\left( H_{v}^{I}-\left( \left( \ln \left\vert C\right\vert \right)
_{,t}u\right) \frac{1}{C}u^{,t}-\frac{1}{2C}\left( \ln \left\vert
C\right\vert \right) _{,tt}u^{2}\right) .
\end{equation}%
\begin{equation}
I^{x}\left( X_{1}\right) =\frac{e^{-\frac{t\left( 1+\alpha _{1}\right) }{c}%
}C^{4}}{2}\left( xH_{v}^{I}-\left( c\left( \ln \left\vert C\right\vert
\right) _{,t}u\right) \frac{1}{C}u^{,x}\right) .
\end{equation}%
\begin{equation}
I^{y}\left( X_{1}\right) =\frac{e^{-\frac{t\left( 1+\alpha _{1}\right) }{c}%
}C^{4}}{2}\left( \alpha _{1}yH_{v}^{I}-\left( c\left( \ln \left\vert
C\right\vert \right) _{,t}u\right) \frac{1}{C}u^{,y}\right) .
\end{equation}%
\begin{equation}
I^{z}\left( X_{1}\right) =\frac{e^{-\frac{t\left( 1+\alpha _{1}\right) }{c}%
}C^{4}}{2}\left( -\left( c\left( \ln \left\vert C\right\vert \right)
_{,t}u\right) \frac{1}{N}u^{,z}\right) .
\end{equation}%
where
\begin{equation}
H_{v}^{I}=\left( \left( -\frac{1}{C^{2}}u_{,t}^{2}+\frac{e^{\frac{2t}{c}}}{%
C^{2}}u_{,x}^{2}+\frac{e^{\frac{2\alpha _{1}}{c}t}}{C^{2}}u_{,y}^{2}+\frac{1%
}{C^{2}}u_{,z}^{2}\right) -V_{v}^{I}\left( t,x,y,z\right) u^{2}\right)
\end{equation}

For the potential function\ $V_{vi}^{I}\left( t,x,y,z\right) $ the resulting
conservation law is derived%
\begin{equation}
I^{t}=I^{t}\left( X_{1}\right) ,
\end{equation}%
\begin{equation}
I^{x}=I^{x}\left( X_{1}\right) +\alpha I^{x}\left( \xi _{I}^{1}\right) ~,
\end{equation}%
\begin{equation}
I^{y}=I^{y}\left( X_{1}\right) +\beta I^{y}\left( \xi _{I}^{1}\right) ~,
\end{equation}%
\begin{equation}
I^{z}=I^{z}\left( X_{1}\right) +\gamma I^{z}\left( \xi _{I}^{1}\right) ~,
\end{equation}%
for $N\left( t\right) =C\left( t\right) ,~A\left( t\right) =e^{-\frac{t}{c}%
}C\left( t\right) ~$and $B\left( t\right) =e^{-\frac{\alpha _{1}}{c}%
t}C\left( t\right) $ with potential function $V_{vi}^{I}\left(
t,x,y,z\right) $

For the potential function$V_{vii}^{I}$ where $X_{2}$ is the generator of
the Lie symmetry vector the resulting Noetherian conservation law is
\begin{equation}
I^{t}\left( X_{2}\right) =\frac{C^{4}}{2}t^{2-\frac{1+\alpha _{1}}{\alpha
_{2}}}\left( \alpha _{2}tH_{vii}^{I}-\alpha _{2}\left( \left( \ln \left\vert
C\right\vert \right) _{,t}u\right) \frac{1}{C}u^{,t}-\frac{1}{2C}\alpha _{2}%
\left[ 1+t(\ln \left\vert C\right\vert )_{,t}\right] _{,t}u^{2}\right) .
\end{equation}%
\begin{equation}
I^{x}\left( X_{2}\right) =\frac{C^{4}}{2}t^{2-\frac{1+\alpha _{1}}{\alpha
_{2}}}\left( xH_{vii}^{I}-\left( \alpha _{2}\left[ 1+t(\ln \left\vert
C\right\vert )_{,t}\right] u\right) \frac{1}{C}u^{,x}\right) .
\end{equation}%
\begin{equation}
I^{y}\left( X_{2}\right) =\frac{C^{4}}{2}t^{2-\frac{1+\alpha _{1}}{\alpha
_{2}}}\left( \alpha _{1}yH_{vii}^{I}-\left( \alpha _{2}\left[ 1+t(\ln
\left\vert C\right\vert )_{,t}\right] u\right) \frac{1}{C}u^{,y}\right) .
\end{equation}%
\begin{equation}
I^{z}\left( X_{2}\right) =\frac{C^{4}}{2}t^{2-\frac{1+\alpha _{1}}{\alpha
_{2}}}\left( \alpha _{2}zH_{vii}^{I}-\left( \alpha _{2}\left[ 1+t(\ln
\left\vert C\right\vert )_{,t}\right] u\right) \frac{1}{N}u^{,z}\right) .
\end{equation}%
where
\begin{equation}
H_{vii}^{I}=\left( \left( -\frac{1}{C^{2}}u_{,t}^{2}+\frac{1}{C^{2}t^{2-%
\frac{2}{\alpha _{2}}}}u_{,x}^{2}+\frac{1}{C^{2}t^{2-\frac{2\alpha _{1}}{%
\alpha _{2}}}}u_{,y}^{2}+\frac{1}{C^{2}}u_{,z}^{2}\right) -V_{vii}^{I}\left(
t,x,y,z\right) u^{2}\right) .
\end{equation}

Finally, for the potential function $V_{viii}^{I}\left( t,x,y,z\right) $ the
conservation law is
\begin{equation}
I^{t}=I^{t}\left( X_{2}\right) ,
\end{equation}%
\begin{equation}
I^{x}=I^{x}\left( X_{2}\right) +\alpha I^{x}\left( \xi _{I}^{1}\right) ~,
\end{equation}%
\begin{equation}
I^{y}=I^{y}\left( X_{2}\right) +\beta I^{y}\left( \xi _{I}^{1}\right) ~,
\end{equation}%
\begin{equation}
I^{z}=I^{z}\left( X_{2}\right) +\gamma I^{z}\left( \xi _{I}^{1}\right) ~,
\end{equation}%
for $N\left( t\right) =C\left( t\right) ,~A\left( t\right) =t^{1-\frac{1}{%
\alpha _{2}}}C\left( t\right) $ ,$~B\left( t\right) =t^{1-\frac{\alpha _{1}}{%
\alpha _{2}}}C\left( t\right) $ and potential function $V_{viii}^{I}\left(
t,x,y,z\right) $

\subsection{Bianchi III}

In the Bianchi III geometry, the Klein-Gordon equation reads
\begin{equation}
\left( -\frac{u_{,tt}}{N^{2}}+\frac{u_{,xx}}{A^{2}}+\frac{u_{,yy}}{B^{2}}%
+e^{2x}\frac{u_{,zz}}{C^{2}}\right) +\frac{1}{N^{2}}\left( \frac{N_{.t}}{N}%
u_{,t}-\frac{A_{,t}}{A}u_{,x}-\frac{C_{,t}}{C}u_{,y}-\frac{C_{,t}}{C}%
u_{,z}\right) -\frac{1}{A^{4}}u_{,x}-\frac{V\left( t,x,y,z\right) }{%
A^{2}B^{2}C}u^{2}=0.  \label{kg.02}
\end{equation}

The three KVs of the Bianchi III spacetime are \ $\xi _{III}^{1}=\partial
_{x}+z\partial _{z},~\xi _{III}^{2}=\partial _{y}$ and $\xi
_{III}^{3}=\partial _{z}$. Hence, (i) $\xi _{III}^{1}$ is a Lie symmetry for
equation (\ref{kg.02}) when $V_{i}^{III}\left( t,x,y,z\right) =V\left(
t,y,ze^{-x}\right) $; (ii) $\xi _{III}^{2}$ is a Lie symmetry for $%
V_{ii}^{III}\left( t,x,y,z\right) =V\left( t,x,z\right) $; (iii) $\xi
_{III}^{3}$ is a Lie symmetry when $V_{iii}^{III}\left( t,x,y,z\right)
=V\left( t,x,y\right) $; (iv) $\alpha \xi _{III}^{1}+\beta \xi
_{III}^{2}+\gamma \xi _{III}^{3}$ is a Lie symmetry when $V_{iv}^{III}\left(
t,x,y,z\right) =V\left( t,y-\frac{\beta }{\alpha }x,\left( z+\frac{\gamma }{%
\alpha }\right) e^{-x}\right) $.

For the line (\ref{ln.04}) where the Bianchi III spacetime admits the
additional CKV $X_{3}$, it follows that the vector field $X=X_{3}-\left(
\frac{2}{m}\frac{A_{,t}}{A}+\lambda \right) u\partial _{u}$ is a Lie
symmetry vector for the Klein-Gordon equation (\ref{kg.02}) when
\begin{equation}
V_{v}^{III}\left( t,x,y,z\right) =\frac{1}{A^{2}}e^{-m\lambda t}U\left(
x,ye^{-\frac{m}{2}t},ze^{-\frac{m\lambda }{2}t}\right) +\frac{m\left(
\lambda -1\right) A_{,t}+2A_{,tt}}{2A^{3}}\text{.}  \label{kg.02a}
\end{equation}

Hence, the vector field $X=X_{3}+\alpha \xi _{III}^{1}+\beta \xi
_{III}^{2}+\gamma \xi _{III}^{3}-\left( \frac{2}{m}\frac{A_{,t}}{A}+\lambda
\right) u\partial _{u}$ is a Lie symmetry of equation (\ref{kg.02}) for the
potential function%
\begin{equation}
V_{vi}^{III}\left( t,x,y,z\right) =\frac{1}{A^{2}}e^{-m\lambda t}U\left( x-%
\frac{\alpha m}{2}t,\left( y+\beta \right) e^{-\frac{m}{2}t},\left( z+\frac{%
\gamma }{\lambda +\alpha }\right) e^{-\frac{m\left( \lambda +\alpha \right)
}{2}t}\right) +\frac{m\left( \lambda -1\right) A_{,t}+2A_{,tt}}{2A^{3}}.
\end{equation}

\subsubsection{Invariant functions}

We proceed with the derivation of the invariant functions provided by each
case for the above potential functions. For $V_{i}^{III}\left(
t,x,y,z\right) $ the Lie invariants are $\left\{ t,y,ze^{-x},u\right\} $,
for $V_{ii}^{III}\left( t,x,y,z\right) $ and the Lie symmetry vector $\xi
_{III}^{2}$ we determine the Lie invariants $\left\{ t,x,z,u\right\} $ while
from $\xi _{III}^{3}$ for the potential $V_{iii}^{III}\left( t,x,y,z\right) $
the Lie invariants are $\left\{ t,x,y,u\right\} $. Similarly, for the
generic vector field $\alpha \xi _{III}^{1}+\beta \xi _{III}^{2}+\gamma \xi
_{III}^{3}$ and potential $V_{iv}^{III}\left( t,x,y,z\right) $ the
corresponding Lie invariants are $\left( t,y-\frac{\beta }{\alpha }x,\left(
z+\frac{\gamma }{\alpha }\right) e^{-x},u\right) $.

In the case where the proper CKV produces a Lie symmetry, then for the
Klein-Gordon equation (\ref{kg.02}) with potential function $%
V_{v}^{III}\left( t,x,y,z\right) $ the Lie invariants are \ $\left\{ x,ye^{-%
\frac{m}{2}t},ze^{-\frac{m\lambda }{2}t},e^{m\lambda t}u\right\} $, while
for the potential function $V_{v}^{III}\left( t,x,y,z\right) $ the resulting
Lie invariants are $\left\{ x-\frac{\alpha m}{2}t,\left( y+\beta \right) e^{-%
\frac{m}{2}t},\left( z+\frac{\gamma }{\lambda +\alpha }\right) e^{-\frac{%
m\left( \lambda +\alpha \right) }{2}t},e^{m\lambda t}u\right\} $.

\subsubsection{Conservation laws}

For the Noetherian conservation laws for the Klein-Gordon equation (\ref%
{kg.02}) it follows that for $V_{i}^{III}\left( t,x,y,z\right) $ it follows
\begin{equation}
I^{x}\left( \xi _{III}^{1}\right) =\frac{NABC}{2}e^{-x}\left( \left( \left( -%
\frac{1}{N^{2}}u_{,t}^{2}+\frac{1}{A^{2}}u_{,x}^{2}+\frac{1}{B^{2}}%
u_{,y}^{2}+\frac{e^{2x}}{C^{2}}u_{,z}^{2}\right) -V_{i}^{III}\left(
t,x,y,z\right) u^{2}\right) \right)
\end{equation}%
\begin{equation}
I^{z}\left( \xi _{III}^{1}\right) =\frac{NABC}{2}e^{-x}\left( z\left( \left(
-\frac{1}{N^{2}}u_{,t}^{2}+\frac{1}{A^{2}}u_{,x}^{2}+\frac{1}{B^{2}}%
u_{,y}^{2}+\frac{e^{2x}}{C^{2}}u_{,z}^{2}\right) -V_{i}^{III}\left(
t,x,y,z\right) u^{2}\right) \right)
\end{equation}%
\begin{equation}
I^{t}\left( \xi _{III}^{1}\right) =0\text{ and }I^{y}\left( \xi
_{III}^{1}\right) =0\text{.}
\end{equation}

For the potential function $V_{ii}^{III}\left( t,x,y,z\right) $ the
Noetherian conservation law has the following components%
\begin{equation}
I^{y}\left( \xi _{III}^{2}\right) =\frac{NABC}{2}e^{-x}\left( \left( \left( -%
\frac{1}{N^{2}}u_{,t}^{2}+\frac{1}{A^{2}}u_{,x}^{2}+\frac{1}{B^{2}}%
u_{,y}^{2}+\frac{e^{2x}}{C^{2}}u_{,z}^{2}\right) -V_{ii}^{III}\left(
t,x,y,z\right) u^{2}\right) \right)
\end{equation}%
\begin{equation}
I^{t}\left( \xi _{III}^{2}\right) =0\text{ ,~}I^{x}\left( \xi
_{III}^{2}\right) =0~\text{and }I^{z}\left( \xi _{III}^{2}\right) =0\text{.}
\end{equation}

Similarly, for $V_{iii}^{III}\left( t,x,y,z\right) $ we determine the
conservation law%
\begin{equation}
I^{z}\left( \xi _{III}^{3}\right) =\frac{NABC}{2}e^{-x}\left( \left( \left( -%
\frac{1}{N^{2}}u_{,t}^{2}+\frac{1}{A^{2}}u_{,x}^{2}+\frac{1}{B^{2}}%
u_{,y}^{2}+\frac{e^{2x}}{C^{2}}u_{,z}^{2}\right) -V_{ii}^{III}\left(
t,x,y,z\right) u^{2}\right) \right)
\end{equation}%
\begin{equation}
I^{t}\left( \xi _{III}^{3}\right) =0\text{ ,~}I^{x}\left( \xi
_{III}^{3}\right) =0~\text{and }I^{y}\left( \xi _{III}^{3}\right) =0\text{.}
\end{equation}

For the Klein-Gordon equation with potential $V_{iv}^{III}\left(
t,x,y,z\right) $ the conservation law has the components%
\begin{equation}
I^{x}\left( \alpha \xi _{III}^{1}+\beta \xi _{III}^{2}+\gamma \xi
_{III}^{3}\right) =\alpha I^{x}\left( \xi _{III}^{1}\right) ~,
\end{equation}%
\begin{equation}
I^{y}\left( \alpha \xi _{III}^{1}+\beta \xi _{III}^{2}+\gamma \xi
_{III}^{3}\right) =\beta I^{y}\left( \xi _{III}^{2}\right) ~,
\end{equation}%
\begin{equation}
I^{z}\left( \alpha \xi _{III}^{1}+\beta \xi _{III}^{2}+\gamma \xi
_{III}^{3}\right) =\alpha I^{z}\left( \xi _{III}^{1}\right) +\gamma
I^{z}\left( \xi _{III}^{3}\right) ,
\end{equation}%
and%
\begin{equation}
I^{t}\left( \alpha \xi _{III}^{1}+\beta \xi _{III}^{2}+\gamma \xi
_{III}^{3}\right) =0,
\end{equation}%
with potential function $V_{iv}^{III}\left( t,x,y,z\right) $.

Moreover, for $V_{v}^{III}\left( t,x,y,z\right) $ the conservation law has
the following components%
\begin{equation}
I^{t}\left( X_{3}\right) =\frac{\bar{A}^{4}e^{\frac{\left( 3m\lambda
-1\right) }{2}t}e^{-x}}{2}\left( \frac{2}{m}H_{V}^{III}-\left( \left( \frac{2%
}{m}\frac{\bar{A}_{,t}}{\bar{A}}+\lambda \right) u\right) \frac{1}{A}u^{,t}-%
\frac{1}{2\bar{A}e^{\frac{m}{2}\lambda t}}\left( \frac{2}{m}\frac{\bar{A}%
_{,t}}{\bar{A}}+\lambda \right) _{t}u^{2}\right) .
\end{equation}%
\begin{equation}
I^{x}\left( X_{3}\right) =\frac{\bar{A}^{4}e^{\frac{\left( 3m\lambda
-1\right) }{2}t}e^{-x}}{2}\left( -\left( \left( \frac{2}{m}\frac{\bar{A}_{,t}%
}{\bar{A}}+\lambda \right) u\right) \frac{1}{\bar{A}e^{\frac{1}{2}\lambda t}}%
u^{,x}\right) .
\end{equation}%
\begin{equation}
I^{y}\left( X_{3}\right) =\frac{\bar{A}^{4}e^{\frac{\left( 3m\lambda
-1\right) }{2}t}e^{-x}}{2}\left( yH_{V}^{III}-\left( \left( \frac{2}{m}\frac{%
A_{,t}}{A}+\lambda \right) u\right) \frac{1}{\bar{A}e^{m\left( \lambda
-1\right) t}}u^{,y}\right) .
\end{equation}%
\begin{equation}
I^{z}\left( X_{3}\right) =\frac{\bar{A}^{4}e^{\frac{\left( 3m\lambda
-1\right) }{2}t}e^{-x}}{2}\left( \lambda zH_{V}^{III}-\left( \left( \frac{2}{%
m}\frac{\bar{A}_{,t}}{\bar{A}}+\lambda \right) u\right) \frac{1}{\bar{A}%
e^{-x}}u^{,z}\right) .
\end{equation}%
where now
\begin{equation*}
H_{V}^{III}=\left( \left( -\frac{1}{\bar{A}^{2}e^{m\lambda t}}u_{,t}^{2}+%
\frac{1}{\bar{A}^{2}}u_{,x}^{2}+\frac{1}{\bar{A}^{2}e^{m\left( \lambda
-1\right) t}}u_{,y}^{2}+\frac{e^{2x}}{C^{2}}u_{,z}^{2}\right)
-V_{v}^{III}\left( t,x,y,z\right) u^{2}\right)
\end{equation*}

Finally, for the potential $V_{vi}^{III}\left( t,x,y,z\right) $ the
conservation law for the Klein-Gordon equation (\ref{kg.02}) related to the
generic symmetry vector $X_{3}+\alpha \xi _{III}^{1}+\beta \xi
_{III}^{2}+\gamma \xi _{III}^{3}-\left( \frac{2}{m}\frac{A_{,t}}{A}+\lambda
\right) u\partial _{u}$ has the following components%
\begin{eqnarray}
I^{t} &=&I^{t}\left( X_{3}\right) ~, \\
I^{x} &=&I^{x}\left( X_{3}\right) +\alpha I^{x}\left( \xi _{III}^{1}\right)
~, \\
I^{y} &=&I^{y}\left( X_{3}\right) +\beta I^{y}\left( \xi _{III}^{2}\right) ~,
\\
I^{z} &=&I^{z}\left( X_{3}\right) +\alpha I^{z}\left( \xi _{III}^{1}\right)
+\gamma I^{z}\left( \xi _{III}^{3}\right) ~,
\end{eqnarray}%
with $N=\bar{A}\left( t\right) e^{\frac{m}{2}\lambda t}$, $B=\bar{A}\left(
t\right) e^{\frac{m}{2}\left( \lambda -1\right) t}$ , $C\left( t\right) =%
\bar{A}\left( t\right) $ and $A\left( t\right) =e^{m\lambda t}\bar{A}\left(
t\right) $ and potential function $V_{vii}^{III}\left( t,x,y,z\right) .$

\subsection{Bianchi V}

For the Bianchi V spacetime the Klein-Gordon equation is
\begin{equation}
\left( -\frac{u_{,tt}}{N^{2}}+\frac{u_{,xx}}{A^{2}}+e^{-2x}\left( \frac{%
u_{,yy}}{B^{2}}+\frac{u_{,zz}}{C^{2}}\right) \right) +\frac{1}{N^{2}}\left(
\frac{N_{.t}}{N}u_{,t}-\frac{A_{,t}}{A}u_{,x}-\frac{C_{,t}}{C}u_{,y}-\frac{%
C_{,t}}{C}u_{,z}\right) +\frac{2}{A^{4}}u_{,x}-\frac{V\left( t,x,y,z\right)
}{A^{2}B^{2}C}u^{2}=0.  \label{kg.03}
\end{equation}%
The KVs of the Bianchi V spacetime are $\xi _{V}^{1}=\partial _{x}-y\partial
_{z}-z\partial _{z}$,~$\xi _{V}^{2}=\partial _{y}$ and $\xi
_{V}^{3}=\partial _{z}$.

Therefore, from Theorem \ref{KG} we find that (i) $\xi _{III}^{1}$ is a Lie
symmetry for the Klein-Gordon equation (\ref{kg.03}) when $V_{i}^{V}\left(
t,x,y,z\right) =V\left( t,e^{x}y,e^{x}z\right) $; (ii) $\xi _{V}^{2}$ is a
Lie symmetry for $V_{ii}^{V}\left( t,x,y,z\right) =V\left( t,x,z\right) $;
(iii) $\xi _{V}^{3}$ is a Lie symmetry when $V_{iii}^{V}\left(
t,x,y,z\right) =V\left( t,x,y\right) $; (iv) $\alpha \xi _{V}^{1}+\beta \xi
_{V}^{2}+\gamma \xi _{V}^{3}$ is a Lie symmetry when $V_{iv}^{V}\left(
t,x,y,z\right) =V\left( t,\left( y-\frac{\beta }{\alpha }\right)
e^{x},\left( z-\frac{^{\gamma }}{\alpha }\right) e^{x}\right) $.

Finally, for the line element (\ref{ln.05}) the vector field $X=X_{3}-\left(
\frac{2}{m}\frac{A_{,t}}{A}+\lambda \right) u\partial _{u}$ is a Lie
symmetry for the Klein-Gordon equation when the potential is of the form of
function (\ref{kg.02a}), while $X=X_{3}+\alpha \xi _{V}^{1}+\beta \xi
_{V}^{2}+\gamma \xi _{V}^{3}-\left( \frac{2}{m}\frac{A_{,t}}{A}+\lambda
\right) u\partial _{u}$ is a Lie symmetry when
\begin{eqnarray}
V_{v}^{V}\left( t,x,y,z\right) &=&\frac{1}{A^{2}}e^{-m\lambda t}U\left( x-%
\frac{\alpha m}{2}t,\left( y+\frac{\beta }{\alpha -1}\right) e^{\frac{%
m\left( \alpha -1\right) }{2}t},\left( z+\frac{\gamma }{\alpha -\lambda }%
\right) e^{-\frac{m\left( \alpha -\lambda \right) }{2}t}\right)  \notag \\
&&~~+\frac{m\left( \lambda -1\right) A_{,t}+2A_{,tt}}{2A^{3}}.
\end{eqnarray}

\subsubsection{Invariant functions}

As previously, we determine the Lie invariants related to the admitted
symmetry vectors for each potential functional. Indeed, for the potential $%
V_{i}^{V}\left( t,x,y,z\right) $ the invariant functions related to the Lie
symmetry $\xi _{III}^{1}$ are $\left\{ t,e^{x}y,e^{x}z,u\right\} $. For the
potential function $V_{ii}^{V}\left( t,x,y,z\right) $ we determine the
invariants $\left\{ t,x,z,u\right\} $ while for $V_{iii}^{V}\left(
t,x,y,z\right) $ the invariants are $\left\{ t,x,y,u\right\} $. Moreover,
for $V_{iv}^{V}\left( t,x,y,z\right) $ the corresponding invariant functions
related to the generic vector field $\alpha \xi _{V}^{1}+\beta \xi
_{V}^{2}+\gamma \xi _{V}^{3}$ are $\left\{ t,\left( y-\frac{\beta }{\alpha }%
\right) e^{x},\left( z-\frac{^{\gamma }}{\alpha }\right) e^{x},u\right\} $.

Finally, for the remaining cases where the proper CKV generates Lie
symmetries, it follows that for potential $V_{v}^{III}\left( t,x,y,z\right) $
the Lie invariants are $\left\{ x,ye^{-\frac{m}{2}t},ze^{-\frac{m\lambda }{2}%
t},e^{m\lambda t}u\right\} $ while for the potential $V_{v}^{V}\left(
t,x,y,z\right) $ and the Lie symmetry $X_{3}+\alpha \xi _{V}^{1}+\beta \xi
_{V}^{2}+\gamma \xi _{V}^{3}-\left( \frac{2}{m}\frac{A_{,t}}{A}+\lambda
\right) u\partial _{u}$ the corresponding Lie invariants are $\left\{ x-%
\frac{\alpha m}{2}t,\left( y+\frac{\beta }{\alpha -1}\right) e^{\frac{%
m\left( \alpha -1\right) }{2}t},\left( z+\frac{\gamma }{\alpha -\lambda }%
\right) e^{-\frac{m\left( \alpha -\lambda \right) }{2}t},e^{m\lambda
t}u\right\} $.

We proceed with the derivation of the conservation laws.

\subsubsection{Conservation laws}

The conservation law for the Klein-Gordon equation (\ref{kg.03}) and
potential function $V_{i}^{V}\left( t,x,y,z\right) $ has the following
components

\begin{equation}
I^{x}\left( \xi _{V}^{1}\right) =\frac{NABC}{2}e^{x}\left( \left( \left( -%
\frac{1}{N^{2}}u_{,t}^{2}+\frac{1}{A^{2}}u_{,x}^{2}+\frac{e^{-x}}{B^{2}}%
u_{,y}^{2}+\frac{e^{-x}}{C^{2}}u_{,z}^{2}\right) -V_{i}^{V}\left(
t,x,y,z\right) u^{2}\right) \right) ,
\end{equation}%
\begin{equation}
I^{y}\left( \xi _{V}^{1}\right) =\frac{NABC}{2}e^{x}\left( -y\left( \left( -%
\frac{1}{N^{2}}u_{,t}^{2}+\frac{1}{A^{2}}u_{,x}^{2}+\frac{e^{-x}}{B^{2}}%
u_{,y}^{2}+\frac{e^{-x}}{C^{2}}u_{,z}^{2}\right) -V_{i}^{V}\left(
t,x,y,z\right) u^{2}\right) \right) ,
\end{equation}%
\begin{equation}
I^{z}\left( \xi _{V}^{1}\right) =\frac{NABC}{2}e^{x}\left( -z\left( \left( -%
\frac{1}{N^{2}}u_{,t}^{2}+\frac{1}{A^{2}}u_{,x}^{2}+\frac{e^{-x}}{B^{2}}%
u_{,y}^{2}+\frac{e^{-x}}{C^{2}}u_{,z}^{2}\right) -V_{i}^{V}\left(
t,x,y,z\right) u^{2}\right) \right) ,
\end{equation}%
and%
\begin{equation}
I^{t}\left( \xi _{V}^{1}\right) =0\text{ .}
\end{equation}

For $V_{ii}^{V}\left( t,x,y,z\right) $ it follows%
\begin{equation}
I^{y}\left( \xi _{V}^{2}\right) =\frac{NABC}{2}e^{x}\left( \left( \left( -%
\frac{1}{N^{2}}u_{,t}^{2}+\frac{1}{A^{2}}u_{,x}^{2}+\frac{e^{-x}}{B^{2}}%
u_{,y}^{2}+\frac{e^{-x}}{C^{2}}u_{,z}^{2}\right) -V_{ii}^{V}\left(
t,x,y,z\right) u^{2}\right) \right) ,
\end{equation}%
\begin{equation}
I^{t}\left( \xi _{V}^{2}\right) =0\text{ ,~}I^{x}\left( \xi _{V}^{2}\right)
=0\text{ and }I^{z}\left( \xi _{V}^{2}\right) =0\text{ }.
\end{equation}

For $V_{iii}^{V}\left( t,x,y,z\right) $ we find
\begin{equation}
I^{z}\left( \xi _{V}^{3}\right) =\frac{NABC}{2}e^{x}\left( \left( \left( -%
\frac{1}{N^{2}}u_{,t}^{2}+\frac{1}{A^{2}}u_{,x}^{2}+\frac{e^{-x}}{B^{2}}%
u_{,y}^{2}+\frac{e^{-x}}{C^{2}}u_{,z}^{2}\right) -V_{iii}^{V}\left(
t,x,y,z\right) u^{2}\right) \right) ,
\end{equation}%
\begin{equation}
I^{t}\left( \xi _{V}^{3}\right) =0\text{ ,~}I^{x}\left( \xi _{V}^{3}\right)
=0\text{ and }I^{y}\left( \xi _{V}^{3}\right) =0\text{ }.
\end{equation}

For the generic vector field $\alpha \xi _{V}^{1}+\beta \xi _{V}^{2}+\gamma
\xi _{V}^{3}$ and potential $V_{iv}^{V}\left( t,x,y,z\right) $ it follows%
\begin{equation*}
I^{x}\left( \alpha \xi _{V}^{1}+\beta \xi _{V}^{2}+\gamma \xi
_{V}^{3}\right) =\alpha I^{x}\left( \xi _{V}^{1}\right)
\end{equation*}%
\begin{equation*}
I^{y}\left( \alpha \xi _{V}^{1}+\beta \xi _{V}^{2}+\gamma \xi
_{V}^{3}\right) =\alpha I^{y}\left( \xi _{V}^{1}\right) +\beta I^{y}\left(
\xi _{V}^{2}\right)
\end{equation*}%
\begin{equation*}
I^{z}\left( \alpha \xi _{V}^{1}+\beta \xi _{V}^{2}+\gamma \xi
_{V}^{3}\right) =\alpha I^{z}\left( \xi _{V}^{1}\right) +\gamma I^{z}\left(
\xi _{V}^{3}\right)
\end{equation*}%
and
\begin{equation*}
I^{t}\left( \alpha \xi _{V}^{1}+\beta \xi _{V}^{2}+\gamma \xi
_{V}^{3}\right) =0,
\end{equation*}%
with potential function $V_{iv}^{V}\left( t,x,y,z\right) .$

From the Lie symmetry vector $X=X_{3}-\left( \frac{2}{m}\frac{A_{,t}}{A}%
+\lambda \right) u\partial _{u}$ we determine the conservation law%
\begin{equation}
I^{t}\left( X_{3}\right) =\frac{\bar{A}^{4}e^{\frac{\left( 3m\lambda
-1\right) }{2}t}e^{x}}{2}\left( \frac{2}{m}H_{V}^{V}-\left( \left( \frac{2}{m%
}\frac{\bar{A}_{,t}}{\bar{A}}+\lambda \right) u\right) \frac{1}{A}u^{,t}-%
\frac{1}{2\bar{A}e^{\frac{m}{2}\lambda t}}\left( \frac{2}{m}\frac{\bar{A}%
_{,t}}{\bar{A}}+\lambda \right) _{t}u^{2}\right) .
\end{equation}%
\begin{equation}
I^{x}\left( X_{3}\right) =\frac{\bar{A}^{4}e^{\frac{\left( 3m\lambda
-1\right) }{2}t}e^{x}}{2}\left( -\left( \left( \frac{2}{m}\frac{\bar{A}_{,t}%
}{\bar{A}}+\lambda \right) u\right) \frac{1}{\bar{A}e^{\frac{1}{2}\lambda t}}%
u^{,x}\right) .
\end{equation}%
\begin{equation}
I^{y}\left( X_{3}\right) =\frac{\bar{A}^{4}e^{\frac{\left( 3m\lambda
-1\right) }{2}t}e^{x}}{2}\left( yH_{V}^{V}-\left( \left( \frac{2}{m}\frac{%
A_{,t}}{A}+\lambda \right) u\right) \frac{1}{\bar{A}e^{m\left( \lambda
-1\right) t}e^{x}}u^{,y}\right) .
\end{equation}%
\begin{equation}
I^{z}\left( X_{3}\right) =\frac{\bar{A}^{4}e^{\frac{\left( 3m\lambda
-1\right) }{2}t}e^{x}}{2}\left( \lambda zH_{V}^{V}-\left( \left( \frac{2}{m}%
\frac{\bar{A}_{,t}}{\bar{A}}+\lambda \right) u\right) \frac{1}{\bar{A}e^{x}}%
u^{,z}\right) .
\end{equation}%
in which%
\begin{equation*}
H_{V}^{V}=\left( \left( -\frac{1}{\bar{A}^{2}e^{m\lambda t}}u_{,t}^{2}+\frac{%
1}{\bar{A}^{2}}u_{,x}^{2}+\frac{e^{-x}}{\bar{A}^{2}e^{m\left( \lambda
-1\right) t}}u_{,y}^{2}+\frac{e^{-x}}{C^{2}}u_{,z}^{2}\right)
-V_{v}^{III}\left( t,x,y,z\right) u^{2}\right) .
\end{equation*}

Finally, the Klein-Gordon equation in the Bianchi V background space with
potential function$~V_{v}^{V}\left( t,x,y,z\right) $ admits the conservation
law with components%
\begin{eqnarray}
I^{t} &=&I^{t}\left( X_{3}\right) ~, \\
I^{x} &=&I^{x}\left( X_{3}\right) +\alpha I^{x}\left( \xi _{V}^{1}\right) ~,
\\
I^{y} &=&I^{y}\left( X_{3}\right) +\alpha I^{y}\left( \xi _{V}^{1}\right)
+\beta I^{y}\left( \xi _{V}^{2}\right) ~, \\
I^{z} &=&I^{z}\left( X_{3}\right) +\alpha I^{z}\left( \xi _{V}^{1}\right)
+\gamma I^{z}\left( \xi _{V}^{3}\right) ~.
\end{eqnarray}

\section{Conclusions}

\label{sec6}

We performed a detailed study for infinitesimal transformations which leave
invariant the Klein-Gordon equation with a non-constant potential function
in curved spacetimes. Specifically, we determined all the admitted Lie and
Noether symmetries for the Klein-Gordon equation. We considered
four-dimensional Riemannian manifolds which describe homogeneous and
anisotropic cosmologies. We wrote the Klein-Gordon equation in the case of
Bianchi I, Bianchi III and Bianchi V spacetimes and we determined all the
unknown functional forms of the potential function where the Klein-Gordon
equations admit non-trivial Lie and Noether symmetries.

We made use of some previous results which relate the infinitesimal
transformations, i.e. the Lie and Noether symmetries, for the Klein-Gordon
equation to the elements of the conformal algebra for the metric tensor of
the Riemannian manifold where the Laplace operator is defined. Thus, we
performed a detailed presentation of the CKVs for the three spacetimes of
our consideration. These spacetimes for arbitrary scale factors have a
three-dimensional conformal algebra which consists of these KVs. However,
for special functions of the scale factors the spacetimes admit a proper
CKV. There are two forms for the line-element of Bianchi I spacetime where a
proper CKV exists, and there is one specific form for the line element of
Bianchi III and one specific line element for the Bianchi V spacetime where
one proper CKV exist.

Thus, for all the specific line elements we present in a systematic way all
the functional forms for the potential for the Klein-Gordon equation where
Lie and Noether symmetries exist. Such an analysis is important in order to
understand the relation of symmetries of differential equations with the
background geometry, as it also shows how symmetries can be derived in a
simple and systematic approach by using tools from differential geometry.
Last but not least, the Noetherian conservation laws can be easily
constructed with the application of Noether's second theorem.

\bigskip

\bigskip

\textbf{Acknowledgments:} The author thanks for the support of Vicerrector%
\'{\i}a de Investigaci\'{o}n y Desarrollo Tecnol\'{o}gico (Vridt) at
Universidad Cat\'{o}lica del Norte through N\'{u}cleo de Investigaci\'{o}n
Geometr\'{\i}a Diferencial y Aplicaciones, Resoluci\'{o}n Vridt No -
096/2022.

\end{document}